\documentclass[twocolumn]{revtex4}
\usepackage{latexsym}
\usepackage{amsmath}
\usepackage{times}
\usepackage{amssymb}
\usepackage{fancyheadings}
\usepackage[T1]{fontenc}
\usepackage[utf8]{inputenc}
\usepackage{url}
\usepackage{hyperref}
\usepackage{color}
\usepackage{graphicx}
 \usepackage{epsfig}

\begin{document}

\title{\Large Geometrical features of time series provide new perspectives on collective fluctuations in driven disordered systems}

\author{Bosiljka Tadi\'{c}\textsuperscript{1}, Miroslav Andjelkovi\'{c}\textsuperscript{2}, Neelima Gupte\textsuperscript{3}}
\affiliation{\textsuperscript{1} Department of Theoretical Physics, Jo\v{z}ef Stefan Institute, Ljubljana, Slovenia\\ \textsuperscript{2} Institute for Nuclear Sciences, Vin\v{c}a, Belgrade, Serbia\\ 
\textsuperscript{3} Indian Institute of Technology Madras, Chennai, India}

\begin{abstract}
Mapping time series onto graphs and the use of graph theory methods opens up the possibility to study the structure of the phase space manifolds underlying the fluctuations of a dynamical variable. Here, we propose to go beyond the standard graph measures and analyze the higher-order structures such as triangles, tetrahedra and higher-order cliques and their complexes occurring in the time-series networks, which are detectable by the algebraic topology methods.  We apply the methodology to the signal of Barkhausen noise accompanying the domain-wall dynamics on the hysteresis loop of disordered ferromagnets driven by the external field. Our analysis demonstrates how the appearance of the complexes with cliques of a high order correlates to the enhanced collective fluctuations in the central part of the hysteresis loop, in contrast to the fractional Gaussian noise fluctuations at the beginning of the loop, which correspond to the graph of a simpler topology. The multifractal analysis of the corresponding segments of these time series confirms that we deal with different types of the stochastic process.
\end{abstract}
%\pacs{12.60.Jv; 12.10.Dm; 98.80.Cq; 11.30.Hv}

\maketitle
\thispagestyle{fancy}

\section{Introduction}
Understanding the emergence of cooperative behavior, as well as the aggregate fluctuations near dynamical phase transitions, constitute challenging problems in the theory of complex dynamical systems. The signatures of the collective dynamics, which are contained in the time series of an observable, are conveniently revealed by methods of time series analysis \cite{TSbook}. Apart from the conventional methods based on the (multi)fractal and information-theory characterization of time series, recently mapping the time series onto analytical graphs has been employed \cite{visibility0,visibility1,visibility2,visibility3,visibility-small14,TSgraphs-turbulence,we-PRE,we-JCSMD,visibility-NG,TSgraphs-Ising}.  The applications of the time-series--graphs duality have revealed new features of the studied dynamical systems. Importantly, it has been recognised that the graph representation and the use of methods of the formal graph theory \cite{BB-book,SD-manual,modularity} and algebraic topology of graphs \cite{JJbook,Qanalysis,BKalg} provide new insights into the nature of aggregate fluctuations by revealing the connection complexity in the phase-space manifolds that underly these fluctuations. In particular,  at the equilibrium phase transition of the Ising spin model the graphs representing the fluctuations of the magnetisation above and below the transition temperature exhibit different structures; the standard graph measures obey a  marked change at the transition point \cite{TSgraphs-Ising}. The situation is more subtle in the case of a nonequilibrium dynamical systems, where the graphs of different structure correspond to altered dynamical regimes, for instance,  across the  turbulent flow \cite{TSgraphs-turbulence}, traffic regimes along the jamming transition \cite{we-PRE}, or the changed collective conduction effects induced by the architecture of the underlying  nanoparticle assembly \cite{we-JCSMD}. In contrast to the equilibrium phase transition \cite{TSgraphs-Ising}, where the transition point is sharp, the changes between the dynamical regimes often take a larger region of the control parameters; consequently, the corresponding networks can not be clearly distinguished by their standard graph measures. Recently, we have proposed to use the algebraic topology methods \cite{Qanalysis} to characterise the differences in the connections among the network's nodes by the presence of the higher-order structures. As it was shown in \cite{we-PRE,we-JCSMD}, different topology measures based on the Q-analysis appear as a sensitive quantifier of the altered dynamical regimes.

Here, we apply this methodology to study the fluctuations of the domain-wall dynamics in driven disordered systems exhibiting the hysteresis loop criticality. The hysteresis phenomena in these systems are suitably described by the zero-temperature random-field Ising model (ZTRFIM) \cite{RFIM-HL,RFIM-EV,RFIM-BT96,RFIM-DD,RFIM-Paco}. The magnetisation jumps corresponding to the increased external field, known as the Barkhausen noise, result from the complex motion of the domain walls that accommodate the spin alignments along the applied field. In this respect, the nature of the noise signal strictly depends on the domain structure of the material, that is, on the strength of the disorder. The disorder is represented by the quenched local magnetic fields \cite{RFIM-HL,RFIM-EV} and, possibly, additional structural defects  \cite{RFIM-BT96,BT-DW}.
Thus, two different regimes can be distinguished. 
The regime of the strong disorder is characterized by small domains and consequently by the appearance of small avalanches of the magnetisation reversal when the field is increased. On the other hand, in the regime with weak disorder avalanches of system size can occur, compatible with large domains. A critical point separates these regimes at a critical value of the random-field disorder $f_{c}$, which resembles an equilibrium phase transition both in three-dimensional and two-dimensional models \cite{RFIM-EV,RFIM-BTUN,RFIM-Djole}. However, the spin dynamics in these reversal processes is different from the equilibrium system, as was shown by the non-perturbative renormalization group \cite{RGnonperturbative}.  Also, it was demonstrated recently \cite{BT-MFR} that the Barkhausen noise in these model systems exhibits a multifractal structure, similar to that seen in many complex dynamical systems driven away from the equilibrium \cite{MFR-examples,mfr-turbulence2013}.
The collective fluctuations may occur near the coercive field in the middle section of the hysteresis loop in a weak disorder regime; in this case, the motion of an extended domain wall can be followed through the random-field obstacles. In contrast, in a strong disorder regime, many domain walls occur and move together thus preventing the depinning transition \cite{BT-DW,BT-MFR}.  

We consider the Barkhausen noise time series for a representative value of the random field disorder below the critical value in the 3-dimensional sample. Selecting the segments of interest on the hysteresis loop, i.e., the initial section and the middle section, we  map the corresponding parts of the time series onto the visibility graphs using the technique of Ref.\ \cite{visibility1,we-PRE}. Then using the techniques of the algebraic topology of graphs, we determine the corresponding topology measures defined in Refs.\ \cite{we-PRE,we-JCSMD} to characterize the higher-order structures in these graphs. Our results show that  shifts in the topology occur,  indicating the dynamical changes in the collective fluctuations as the system is driven along the hysteresis loop. For comparison, we systematically determine the multifractal features of the same time series, using the approach described in Ref.\ \cite{BT-MFR}.

\section{Results}

We simulated the process of the magnetization reversal along the hysteresis loop of the 3-dimensional ZTRFIM with the ferromagnetic spin-spin interaction of strength $J$ and Gaussian distribution of the random fields. Then we analyse the temporal correlations and multifractal properties of the time series of the magnetisation fluctuations by considering different segments of the hysteresis loop. Finally, we map these parts  of the time series onto the graphs (DW-networks) and analyse their structure.

\subsection{Model and simulations of domain wall dynamics \label{sec-RFIM}}
To simulate the magnetisation reversal time series, we use a three-dimensional version of the random-field  Ising model on a cubic $L\times L \times L$ lattice with $L
 = 50$. The quenched disorder is implemented  by the Gaussian random fields $h_i$ with a zero  mean and standard deviation $f$. The system of the ferromagnetically interacting spins is described by the  Hamiltonian
\begin{equation}
\mathcal{H}=-\sum_{<ij>} J_{ij} S_{i} S_{j} -\sum_{i} (h_i+H) S_{i} \equiv -\sum_iS_ih_i^{loc}(t)
\label{eq-Ham}
\end{equation}
where the sum $<ij>$ is over nearest neighbours. Here, to use dimensionless units, we set $J_{ij} =J=1$ to be a constant interaction between Ising spins $S_{i} = \pm 1$.  The system is driven along the hysteresis loop by the external magnetic field $H(t)$.
 In the zero-temperature dynamics, the spin states are determined according to the values of the local fields $h_i^{loc}(t)$, which depends on the current value of the external field, the local random field $h_i$, as well as the contributions from the interacting neighbour spins, cf. (\ref{eq-Ham}). 
In the  initial
configuration all the spins are $S_i=-1$  aligned along a large negative field $H(t_0)=-H_{max}$.
Then the field is ramped  in small steps of $R$; after each field increase, the spin system relaxes to minimise the energy (\ref{eq-Ham}) in the current field value. We use parallel updates, where at each time step $t$ the set of local fields $h_i^{loc}(t)$ is computed and then the corresponding spins aligned with them to minimise the energy. The reversed spins can further destabilize their neighbours, such that the process is repeated as long as the system enters a metastable state in
which all the spins are locally stable. The number of the flipped spins at a particular time step $t$ comprises the magnetisation discontinuity $\delta M(t)$ at time $t$ or the data point of the Barkhausen noise signal; the total number of the reversed spins for a given external field represents a magnetisation-reversal avalanche. In the next time step, the external field is increased  $H  \rightarrow H+R$ again and so on until all spins are reversed at $H_{max}$.  
We apply the periodic boundary conditions in all directions. 
For this work, we fix $f=2.3$ what keeps the random fields in the weak  disorder regime and apply small field steps $R=2\times 10^{-3}$ to reduce the driving-rate effects and  
thus permitting the possibility for the depinning transition in the central part of the loop \cite{we-drivingrate2004,DWstochasticity2011}.

\subsection{Temporal and fractal features of the magnetisation fluctuations in segments of the hysteresis loop\label{sec-temporal}}

\begin{figure}[!htb] 
\centering
\includegraphics[width=.95\columnwidth]{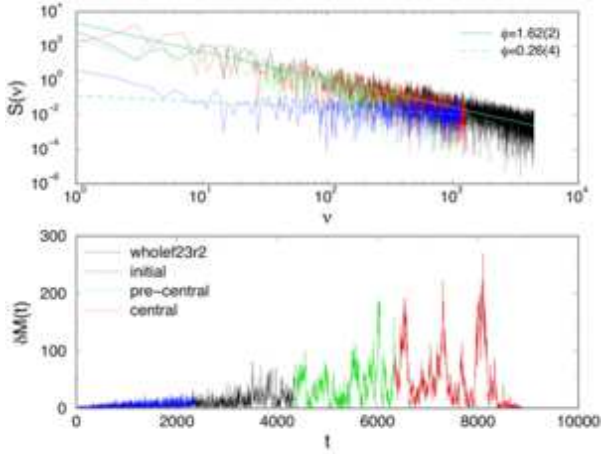}
\caption{Bottom: The time series of the magnetization jumps along the hysteresis loop for the case of the weak disorder. Segments in different parts of the hysteresis loop, each comprising of the $2000$ data points, are indicated by different colors. Top: Power spectrum of the entire time series and the corresponding segments.}
\label {fig1}
\end{figure}

The magnetization reversal time series corresponding to the ascending branch of the hysteresis loop in the case of weak disorder regime is shown in Fig.\ \ref{fig1}. 

Evidently, at the beginning of the process (the initial part of the hysteresis loop), only small fluctuations are present; the fluctuations gradually increase as the field approaches the value where the magnetization changes sign (the coercive field), i.e., the central part of the hysteresis loop. Here, we realize that large fluctuations can occur for each increase of the external field. Through the sequence of large avalanches, the whole system eventually reverses. A detailed description of the avalanches statistics for this case \cite{RFIM-Paco,BT-MFR} shows that some of the avalanches can extend over the entire system, for instance, in a two-dimensional plane, suggesting that a depinning of the domain wall occurred.  Only a few small avalanches are seen at the right end of the signal corresponding to the end of the hysteresis loop. 
For the purpose of this work, we divide the whole time series into segments of the length 2000 data points, starting from the beginning. Different colors in Fig.\ \ref{fig1} indicate these segments.  
The power spectra of the magnetization reversal time series in  Fig.\ \ref{fig1} (top panel) suggest that the long-range temporal correlations occur in this stochastic process. The power-law dependencies of the power spectra suggest a power-law dependence $S(\nu) \sim \nu^{-\phi}$ noise for the whole signal and all segments excluding the initial one. For the initial part of the hysteresis loop, the portion  is shown in blue,  the fluctuations close to the white noise type occur (the slope of the power spectrum is close to zero).

As it was shown in Ref.\ \cite{BT-MFR}, the fluctuations of the magnetisation in the reversal processes represent an example of a complex signal exhibiting multifractal properties. These features of the  Barkhausen noise indicate the dynamical nature of the hysteresis loop criticality; moreover, they suggest that the underlying dynamics of the domain wall motion is a highly stochastic process, which shares some similarity with the turbulence \cite{mfr-turbulence2013} and many other complex dynamical systems driven away from the equilibrium \cite{MFR-examples}. Here, we demonstrate how further differences in the fluctuations of the magnetisation in different segments of the hysteresis loop are detected in the multifractal spectrum of these segments. 

The occurrence of multifractality of a signal, in fact, indicates that some areas of the signal have different scaling behavior; hence, various areas of the time series need to be amplified by the different factor $q\in Re$ to become self-similar to the rest of the signal. 
Technically, applying the detrended multifractal time series analysis \cite{MFR0,MFRA-uspekhi2007,BT-MFR}, we determine the scale-invariance of the fluctuation function  $F_q(n)$ at the varying time interval $n$, which is defined as 
\begin{equation}
F_q(n)=\left\{(1/2N_s)\sum_{\mu=1}^{2N_s} \left[F^2(\mu,n)\right]^{q/2}\right\}^{1/q} \sim n^{H(q)}  \ ,
\label{eq-FqHq}
\end{equation}
where $n\in[2,T_{max}/4]$ starting once from the beginning of the time series and then from the end $T_{max}$.
Here,  $Y(i)=\sum_{k=1}^i(\delta M(k)-\langle \delta M\rangle)$ stands for the integrated original signal $\delta M(k)$, $2N_s=2int(T_{max}/n)$ is the number of devisions of the time series of the length $T_{max}$, and  $F^2(\mu,n)$ is the standrad deviation 
\begin{equation}
 F^2(\mu,n) = \frac{1}{n}\sum_{i=1}^n[Y((\mu-1)n+i)-y_\mu(i)]^2 
\label{eq-F2}
\end{equation}
of the signal from its local trend $y_\mu(i)$ which is determined at each at the segment $\mu=1,2\cdots N_s$.
Given Eqs.\ (\ref{eq-FqHq})-(\ref{eq-F2}), the scaling exponent $H(q)$ of the fluctuation function appears as a generalisation of the standard Hurst exponent, which corresponds to the case $q=2$.

\begin{figure}[htb] 
\centering
\includegraphics[width=0.96\columnwidth]{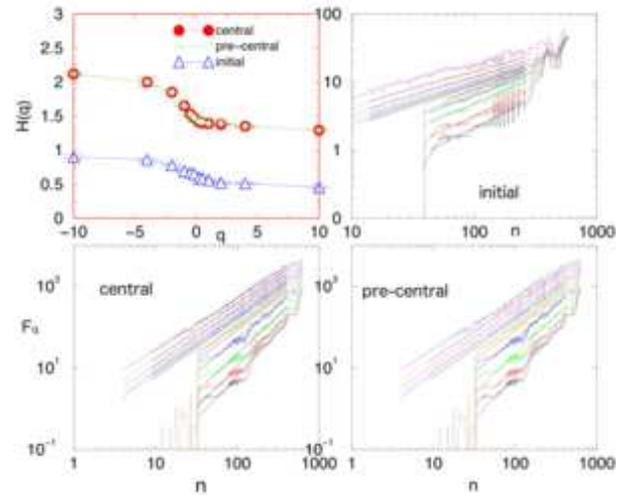}
\caption{The generalized Hurst exponent $H(q)$ plotted against the distortion parameter $q$ and, in  remaining three panels, the corresponding fluctuation function $F_q(n)$ vs. time interval $n$ for the initial, central and  pre-central segment of the magnetization-reversal time series.}
\label{fig2}
\end{figure}

Applying the methodology of the detrended multifractal analysis of time series here we compute the generalised Hurst exponent $H(q)$ for the above-identified segments of the magnetisation fluctuations. 
In Fig.\ \ref{fig2}, we show the fluctuation function $ F_q(n)$  against the time interval $n$, evaluated for a large span of the distortion parameter $q\in [-10,10]$. According to the theory of multifractal time series \cite{MFR0,BT-MFR}, the occurrence of different slopes $H(q)$ of the fluctuation function $ F_q(n)\sim n^{H(q)}$ for different parameters $q$ is a signature of the multifractality. Consequently, we can measure a range of values of the generalised Hurst exponent $H(q)$, as opposed to monofractals, where all exponents are equal to the standard Hurst exponent $H(q)=H(q=2)$. The analysis is performed for all  segments, cf. Fig.\ \ref{fig2}, and the resulting generalised Hurst exponents are shown. These results suggest that the middle parts of the hysteresis loop belong to the type of process known as fractional Brownian motion, where $H(q)>1$ for all $q$. While the fluctuations in the initial segment of the loop represent another kind of the stochastic process, fractional Gaussian noise, for which it remains in the range $H(q)<1$ for all $q$.

\subsection{Topology of DW-networks in different segments  of the hysteresis loop\label{sec-topology}}

\begin{figure}[htb] 
\centering
\includegraphics[width=.92\columnwidth]{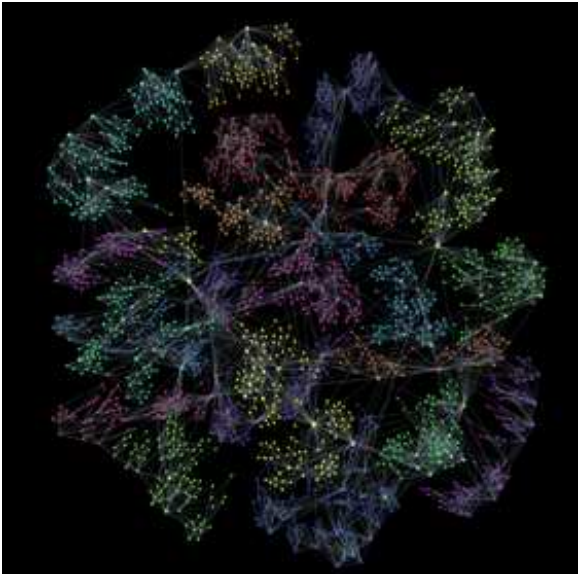}
\caption{The visibility graph representing the time-series from the initial segment of the hysteresis loop. Colours on nodes indicate different communities detected by the maximum-modularity method.}
\label{fig3}
\end{figure}

\begin{figure}[htb] 
\centering
\includegraphics[width=.92\columnwidth]{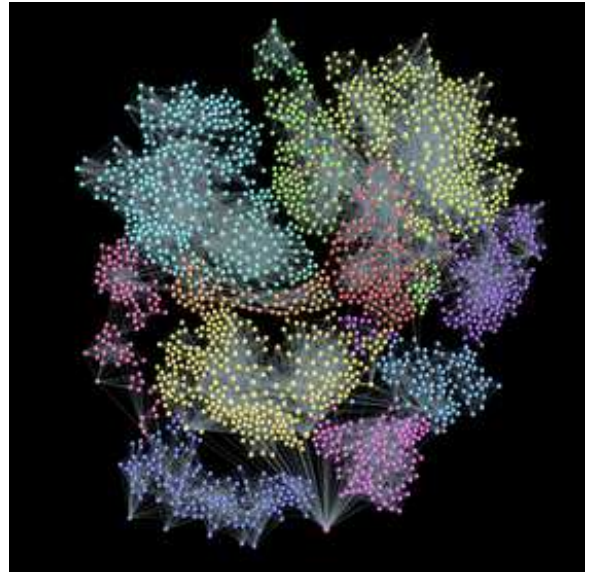}
\caption{The visibility graph of the central loop segment of the magnetization-reversal time series. The community structure of the graph is indicated by different colors of nodes.}
\label{fig4}
\end{figure}

Furthermore, the mapping of these sectors of the time series onto networks by the visibility method \cite{visibility1,we-PRE} results in graphs of different structures. The DW-graphs corresponding to  the initial and central part of the hysteresis loop are shown in Fig.\ \ref{fig3} and Fig.\ \ref{fig4}, respectively.
Visually, the graphs corresponding to the fluctuations in the central and the initial part of the hysteresis loop have different structures. The standard graph-theoretic measures  \cite{BB-book,SD-manual}, the average degree $\langle k\rangle$, the diameter of the network $\langle d\rangle$, the clustering coefficient, and the  path length $\langle\ell \rangle$, are summarised in Table \ref{Table1}. The modularity of these graphs, determined by the modularity maximisation methods \cite{modularity}, manifests in the community structure, as also shown in Figs.\ \ref{fig3} and \ref{fig4}.

\begin{table*}[htb]
\centering
\caption{The standard graph-theoretic measures of the DW-networks in different sectors of the hysteresis loop,  cf. Figs.\ \ref{fig3} and \ref{fig4}}
\label{Table1}
\begin{tabular}{|l|l|l|l|}
\hline
\textbf{Quantifiers}&{\textbf{Initial part}}&\textbf{Pre-central part}&\textbf{Central part}\\
\hline
Average degree $\langle k \rangle$& 7.113&9.85 &12.199\\
Diameter $\langle d \rangle$& 9& 11&11\\
Clustering coefficient & 0.757&0.734 &0.715\\
Modularity&0.906 & 0.877&0.822\\
Average path length $\langle \ell \rangle$&4.979& 4.948&5.108\\
\hline
\end{tabular}
\end{table*}

The major differences, however, are in the organization of the links, which manifests in the presence of the higher-order cliques  and their complexes
in the visibility graphs corresponding to these time series. These differences can be quantified via the topological measures as, for instance, described in \cite{we-PRE}.

The topological measures, or simplicial characteristics defined in \cite{we-PRE}, can be used to analyse any graph or network. Here, by a graph or network, we mean  a collection of nodes interacting via interconnected edges or links as a result of the visibility mapping. We define a clique to be a maximal complete subgraph.
Starting from the adjacency matrix of the graph, the cliques can be identified via the Bron-Kerbosch algorithm\cite{BKalg}. The cliques are recognised as simplexes, and the agglomerates of these cliques are then identified  as the simplicial complexes of the graph. 
Precisely, a clique with $q+1$ nodes or vertices is a $q$ dimensional simplex. If two simplices have $q+1$ nodes in common, they share a $q$ face. A collection of simplices, i.e. the nodes with the shared faces form a simplicial complex. We are interested in the $q-$connectedness of the simplex, as well as in the dimension of the simplicial complex, i.e. the dimension of the largest simplex in the complex. If we can find a sequence of simplices such that each successive pair shares a $q$ face, then all the simplices in this sequence are said to be $q-$connected. Simplices which are $q-$connected, are also connected at all lower levels.  

\begin{figure}[htb] 
\centering
\includegraphics[width=1.0\columnwidth]{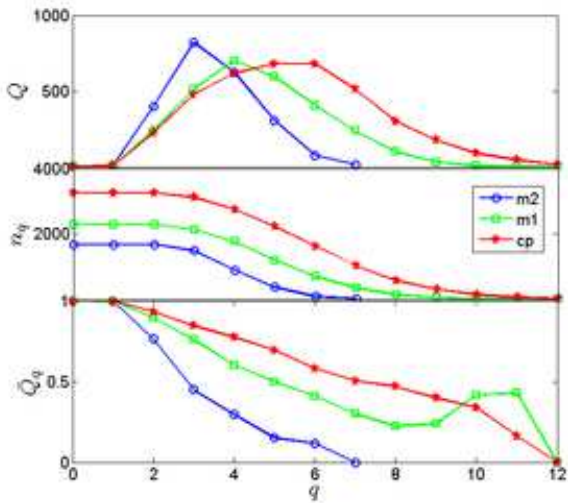}
\caption{The three structure vectors plotted against the topology level $q$ of the graphs corresponding to the three visibility graphs of the magnetization reversal time series in the initial (m2), pre-central (m1) and central (cp) part of the hysteresis loop.}
\label{fig5}
\end{figure}

\begin{figure}[htb] 
\centering
\includegraphics[width=0.98\columnwidth]{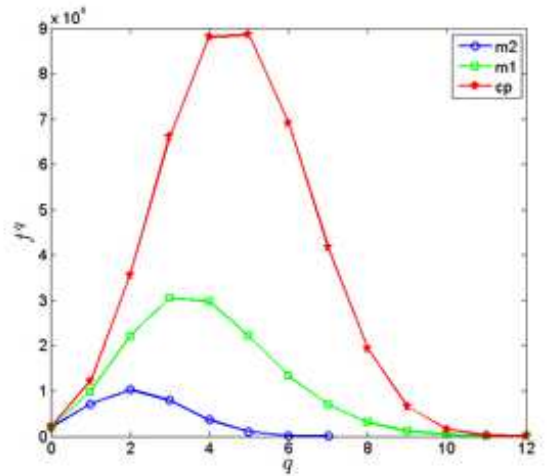}
\caption{The topological response functions $f^q$ against the topology level $q$ for the same graphs as in Fig.\ \ref{fig5}, the abbreviations and the color code apply.}
\label{fig6}
\end{figure}

The occurrence of the higher-order cliques  in the visibility graphs of the time series  is described by three structure vectors and the topological response function introduced in \cite{we-PRE}. 
In particular, the components of the first structure vector $Q_q$ determine the number of the $q$-connected classes at the topology level $q$, while the second structure vector $n_q$ indicates the number of cliques of the order $q$ and larger. Then the combination $\bar{Q}_q=1-Q_q/n_q$ is the third structure vector, which shows how the cliques are connected at each topology level $q$ below the maximal clique. 
(Note that this $q$ should not be confused with the distortion parameter in the multi-fractal analysis of section\ \ref{sec-temporal}).
The three structure vectors as a function of the topology level $q$ for three segments of the time series are given in Fig.\ \ref{fig5}. The corresponding response function $f^q$ are depicted in Fig.\ \ref{fig6}. Remarkably, these figures show that an increasingly more complex topology appears as the system approaches the fluctuations associated with the  domain-wall depinning in the central part of the hysteresis loop. The topological complexity is indicated first by an increase of the $q_{max}$ the largest occurring topology level, as well as the shift of the maximum in the first structure vector as well as the response function  $f^q$ towards the larger topology levels.

Specifically, while the largest clique found in the graph of the initial sector is of the order $q_{max}=7$, the maximum topology level in the inner segments extends to $q_{max}=12$. At the highest level, there are a few separated cliques. The number of cliques below the highest level is gradually increasing with the lower $q$, and it is larger in the most complex topology reaching the maximum at $q=6$ (red line in Fig.\ \ref{fig5} top panel). That is, this graph comprises the largest population of 7-cliques, which are mutually interconnected via a large number of faces (cliques of the lower orders), as indicated by the third structure vector in the bottom panel. Comparably, in the pre-central segment, the graph shows heavily interconnected cliques close to the top level, and considerably lower values of all structure vectors at lower topology levels. Its maximum is at $q=4$, corresponding to 5-cliques. Much simpler topology is found in the graph of the initial segment, where the largest population corresponds to $q=3$, i.e. tetrahedra (cf. Fig.\ref{fig3}); however, they are only weakly interconnected, the corresponding component of the third structure vector at this level is smaller than $0.5$.

The topological response function $f^q$, which is introduced in \cite{we-PRE},  indicates even larger differences between these graphs, as shown in Fig.\ \ref{fig6}.   Recall that $f^q$ measures the number of cliques of the order $q$ but taking into account that some of the cliques are in fact the faces shared by the existing higher-order cliques. Thus, it appears to be a sensitive measure of the aggregations of cliques, which, on the other side, directly reflects the strength of the signal's fluctuations, for instance, compare the graphs in Fig.\ \ref{fig3} and Fig.\ \ref{fig4}. Notably, apart from the lower values for the largest clique found in the graph, $q_{max}$, the response function for the graph in the initial segment of the hysteresis loop has the maximum at $q=2$, i.e., at the level of triangles. The maximum gradually increases and moves towards the higher $q$ values as the topological complexity increases in the case of graphs in the inner parts of the loop. In physical systems, a frequency-dependent dielectric response in the ferroelectric glasses \cite{SGresponse} has a similar behavior, indicating a multiplicity of the time scales in the glassy state. In the present context, removing a triangle is likely to result in the dismantling of an important simplicial complex in the original graph. While in the graph corresponding to the middle segment of the loop  the maximum response can be evoked in a more complex structure by removing a 6-clique that it contains.

\section{Conclusion}
We have considered the time series which are emanating from the magnetization reversal process on the hysteresis loop in weakly disordered ZTRFIM.  The rich dynamical behaviour of the system is dominated by the motion of individual domain walls through the random environment, which is driven by the slowly increasing external magnetic field. While the random obstacles dominate at the low field values at the beginning of the hysteresis loop, the stronger fields in the middle segments of the loop may cause the domain-wall depinning, the collective dynamical behavior resulting in massive fluctuations in the time series.  The power spectrum and multifractal analysis suggest that these fluctuations belong to another class of the fractal stochastic processes, as compared to the initial segment of the loop. By mapping the segments of the time series onto visibility graphs, we have demonstrated how the algebraic topology analysis of these graphs identifies the changes in the dynamical regime as the system is driven along the hysteresis loop.  We have determined several structure vectors and the topology response function to characterize the higher-order structures in these graphs. The computed topology measures are very sensitive in detecting the changes in the topology. Remarkably, they indicate that the systematically more complex topologies occur with the increased collective fluctuations in the inner segments of the loop. These geometrical features of the compound signals cast new light onto the nature of collective fluctuations; they may be useful to detect the enhanced fluctuations in many driven dynamical systems, particularly when the occurrence of a nonequilibrium phase transition is not apparent or other factors may mask the transition.

%%Use section* for acknowledgements
\section*{Acknowledgement}
The authors acknowledge the financial support from the Slovenian
Research Agency (research code funding number P1-0044) and the Projects 
ON174014 by the Ministry
of Education, Science and Technological Development of the Republic of Serbia.
MA and NG thank the hospitality at the Department of Theoretical Physics of the Jo\v zef Stefan Institute.
%Acknowledgements

%%use \balance somewhere in the left column of the last page to balance the two columns in the end page

%%References section

\end{document}